\documentclass[aps,pre,preprint,groupedaddress]{revtex4-1}

\usepackage{amsmath}
\usepackage{amssymb}
\usepackage{graphicx}
\usepackage{siunitx}
\usepackage{xcolor}
\usepackage{physics}

\renewcommand{\vec}[1]{\vectorbold*{#1}}
\newcommand{\uvec}[1]{\hat{\vec{#1}}}

\renewcommand{\ss}[0]{\mathrm{ss}} 
\newcommand{\ms}[0]{\mathrm{ms}}

\newcommand{\sgf}[0]{\mathcal{G}}

\newcommand{\comment}[1]{#1}

\begin{document}

\title{\comment{Dense Plasma Opacity via the Multiple-Scattering Method}}

\author{Nathaniel R. Shaffer}
\email[]{nsha@lle.rochester.edu}
\affiliation{Los Alamos National Laboratory}
\affiliation{University of Rochester Laboratory for Laser Energetics}

\author{Charles E. Starrett}
\affiliation{Los Alamos National Laboratory}

\date{\today}

\begin{abstract}
  The calculation of the optical properties of hot dense plasmas with a model that has self-consistent plasma physics is a grand challenge for high energy density science.
  Here we exploit a recently developed electronic structure model that uses Multiple Scattering Theory to solve the Kohn-Sham density functional theory equations for dense plasmas.
  We calculate opacities in this regime, validate the method, and apply it to recent experimental measurements of opacity for Cr, Ni and Fe.
  Good agreement is found in the quasi-continuum region for Cr and Ni, while the self-consistent plasma physics of the approach cannot explain the observed difference between models and the experiment for Fe.
\end{abstract}

\maketitle

\section{Introduction}
\label{sec:intro}

One of the grand challenges in high energy density physics is a comprehensive model of dense plasma opacity.
Most modern-day plasma opacity models are based on the atomic structure calculations of isolated ions, which are then coupled to external models for plasma density and temperature effects and possibly rate equations for ion population kinetics\cite{FontesJPB2015,IglesiasAJ1996, badnellMNRAS2005,OpusReport}.
Such approaches are well-suited for low-density plasmas, where the isolated-ion picture is a good starting point for the electronic structure and where there exist accurate models for the most important plasma effects such as Stark broadening, ionization potential depression, and continuum lowering.
However, there is a growing corpus of experimental and computational evidence that these models become unreliable for high-density plasmas, taken here to mean densities greater than about 1/100th of solid density\cite{GillPRE2021, ciricostaPRL2012, iglesias2013fluctuations, BaileyNL2015, huPRL2017, massacrierPRR2021, SonPRX1014}.

The essential physical problem is that at high densities, one cannot decouple the atomic physics from the plasma physics.
That is, the electronic structure of individual ions must be considered self-consistently with their neighbors.
Rather than isolated-ion electronic structure, the natural starting point should be a multi-atom electronic structure method that includes a consistent treatment of free electrons.  
In that way, the important temperature and density effects -- Stark broadening and continuum lowering in particular -- are built into the electronic structure from the outset.
One such method that meets this requirement is finite-temperature density functional theory (DFT) \cite{hohenberg1964inhomogeneous,kohn65self, merminPR1965}.

In principle, with DFT all electrons are treated equally, there is no required distinction between free and bound electrons.
Moreover, the electronic structure due to many ions is found at once; there is no need to first consider isolated atoms and try to add plasma effects afterward.
These strengths mean that the plasma physics in a DFT treatment is very high quality.
The method has been used with success for the optical properties and equation of state of dense plasmas for over 20 years \cite{rehrRMP2000, mattern2012real, huPRL2017, mazevet2005simulations, recoulesPRB2009, DesjarlaisPRE2002}. 
Due to computational and practical limitations, the method has been restricted in its applications to systems with degenerate or nearly degenerate electrons (i.e., temperatures similar to, or lower than, the Fermi temperature). 

Recently, however, these limitations have been overcome by adapting Multiple Scattering Theory (MST) to plasma conditions \cite{StarrettPRE2020, LaraiaHEDP2021}.
MST is a blanket term \cite{EbertRPP2011, faulknerBook2018} that covers both the Korringa-Kohn-Rostoker (KKR) method \cite{KorringaP1947,KohnPR1954} as well as the Real-Space Green's Function (RSGF) method \cite{rehrRMP2000, wangRPL1995}.   
Its name derives from the multiple scattering of a wave as it passes though a medium \cite{KorringaPR1994}.  
For electrons, this refers to quantum diffraction.  
Originally developed to model periodic solids \cite{KorringaP1947, KohnPR1954} (KKR), the method was adapted to treat clusters of atoms and molecules (RSGF), with wide ranging use for optical properties \cite{rehrRMP2000, ankudinovPRB1998, PeyrusseHEDP2010, TanPRB2021}.  

In brief, MST solves the electronic structure problem by using a multi-center expansion.
This casts the global electronic structure problem into many local ones, whose solutions are coherently connected to one another.
While not restricted to solving the Kohn-Sham DFT equations (in fact it predates DFT), it has found widespread use for such problems \cite{asato1999full, huhne1998full, EbertRPP2011}. 
In  references \cite{StarrettPRE2020} and \cite{LaraiaHEDP2021} both KKR and RSGF were used to model dense plasmas with DFT and predict their equations of state.
MST has two principal advantages over other methods to solve the DFT equations for plasmas.
First, core electrons are computed consistently, i.e., without pseudo-potentials.
Second, it can reach arbitrarily high temperatures without prohibitive computational scaling.  

In this work, we apply MST to the optical properties of hot dense plasmas\comment{, focusing on conditions that are inaccessible or extremely challenging to existing multi-center methods.}
We compare to existing methods based on DFT and find generally good agreement. 
After this validation, we apply the method to the recent experiments on Cr, Fe and Ni \cite{BaileyNL2015, NagayamaPRL2019}. 
We find good agreement on Cr and Ni for the quasi-continuum (bound-free) region of the spectra, while for Fe, the enigmatic difference between models and the experiment \cite{NagayamaPRL2019} is reinforced by our calculations. 
The comparison with experiment also highlights the weakness of the approach for bound-bound spectra.
\comment{To our knowledge, this is the first multi-center calculation of the opacity for these experiments. 
Moreover, since MST has a consistent treatment of all electrons, effects such as continuum lowering and wave function decoherence due to multiple collisions with ions\cite{LiuNCP2018} are automatically included. 
Ours results therefore imply that the \textit{ad hoc} treatment of these effects in models starting from isolated atom calculations, is not the source of the difference between such models and the experimental results for Fe.}

\section{Theory}
\label{sec:theory}

In a basis of single-particle states, the real part of the optical conductivity is given by the Kubo-Greenwood formula \cite{greenwood1958boltzmann, KuboJPSJ1957}
\begin{equation}
  \label{eq:kg-states}
  \sigma(\omega) = \frac{2\pi e^2\omega}{3\Omega} \sum_{a,b} \abs{\bra{a} \vec r \ket{b}}^2 (f_a - f_b) \delta(\hbar\omega - E_b + E_a)
\end{equation}
where $\hbar\omega$ is the photon energy, $\Omega$ is the system volume, $a$ labels the initial states with energy $E_a$ and Fermi-Dirac occupation $f_a$, and $b$ labels the final states with energy $E_b$ and occupation $f_b$.
Here and throughout, the twofold spin degeneracy of each state has been explicitly factored out of the summations.
Positive $\omega$ corresponds to photoexcitation of state $a$ to $b$, with the Dirac delta representing the conservation of energy.
To evaluate the conductivity using the Green's function formalism, we introduce the one-electron Green's function
\begin{equation}
    \label{eq:lehmann}
  G(\vec r, \vec r'; z) = 2\sum_{a} \frac{\bra{\vec r}\ket{a} \bra{a}\ket{\vec r'}}{z - E_a}
\end{equation}
Here, $z = E + i\Gamma$ is the complex energy with real $E$ and $\Gamma$.
Throughout, we consider $\Gamma \ge 0$, i.e., the retarded Green's function.
As $\Gamma \to 0^+$, the imaginary part of the Green's function is
\begin{equation}
  \Im G(\vec r, \vec r'; E) = -2\pi \sum_{a} \bra{\vec r}\ket{a} \bra{a}\ket{\vec r'} \delta(E - E_a)
\end{equation}
in terms of which the Kubo-Greenwood formula may be rewritten
%
\begin{equation}
  \label{eq:kg-green}
  \sigma(\omega) = \frac{e^2\omega}{6\pi\Omega} \int_\Omega \int_\Omega  \int_{-\infty}^\infty
  (f_1 - f_2) (\vec r\vdot\vec r') 
  \Im G(\vec r,\vec r'; E_1) \Im G(\vec r', \vec r; E_2) \dd{E_1} \dd{\vec r'} \dd{\vec r}
\end{equation}
in which $E_2 = E_1 + \hbar\omega$ and the discrete sums over states have been replaced with an integral over all energies.
The usual form of the Kubo-Greenwood relation, Eq.~\eqref{eq:kg-states}, is recovered using the residue theorem.
The Green's function formulation allows us to make use of an efficient, local representation, called the Real-Space Green's Function (RSGF) method.

The idea of the RSGF method is to partition the domain $\Omega$ into non-overlapping cells, $\{\Omega_n\}$, each containing one special point, $\vec R_n$, called the ``center'' or ``site''.
For crystals, these centers are usually chosen to coincide with the nuclei.
For disordered plasmas, it is necessary to also have cells without nuclei\cite{StarrettPRE2020}.
The $\vec r$ and $\vec r'$ dependence of the Green's function are expressed relative to the center of the cell they lie in
\begin{equation}
  \vec r = \vec r_n + \vec R_n \qquad \vec r' = \vec r'_{n'} + \vec R_{n'}
\end{equation}
so that the Green's function can be expressed piecewise in terms of site-site Green's functions
\begin{equation}
  \label{eq:kkr-piecewise}
  G(\vec r, \vec r'; z) = \begin{cases}
    \quad\vdots \\
    G^{nn'}(\vec r_n, \vec r'_{n'}; z)  \qif \vec r \in \Omega_n , \vec r' \in \Omega_{n'} \\
    \quad\vdots
  \end{cases}
\end{equation}
With this representation of the Green's function, the integrals over all space in Eq.~\eqref{eq:kg-green} can be decomposed into integrals over the cells
\begin{equation}
\begin{split}
  \label{eq:kg-kkr}
  \sigma(\omega) &= \frac{e^2\omega}{6\pi\Omega} \sum_{nn'} \int_{\Omega_n} \dd{\vec r_n} \int_{\Omega_{n'}} \dd{\vec r'_{n'}} \int_{-\infty}^\infty \dd{E_1} \\
  &\times (\vec r_n + \vec R_n)\vdot(\vec r'_{n'} + \vec R_{n'})(f_1 - f_2) \\
  &\times\Im G^{nn'}(\vec r_n, \vec r'_{n'}; E_1) \Im G^{n'n}(\vec r'_{n'}, \vec r_n; E_2) 
\end{split}
\end{equation}
The advantage of this representation is one can expand $G^{nn'}$ in spherical harmonics about the site centers $\vec R_n$ and $\vec R_{n'}$.
This expansion is only conditionally convergent for arbitrarily shaped cells\cite{ZellerJPCM2013,GonisButlerBook}, but this issue is mitigated if the cells are nearly spherical, which is one reason for introducing non-nuclear expansion centers.

The general spherical harmonic expansion of the site-site Green's function has two terms, $G^{nn'} = G^{nn'}_\ss + G^{nn'}_\ms$, these being (with energy arguments suppressed) a single-site part
\begin{equation}
   \label{eq:g-ss}
   G^{nn'}_\ss(\vec r_n, \vec r'_{n'}) = -2\pi i \delta_{nn'} \sum_{L} R_L^n(r_<) H_L^n(r_>) Y_L(\uvec r_n) Y^*_L(\uvec r'_n)
\end{equation}
and a multiple-scattering part
\begin{equation}
   \label{eq:g-ms}
   G^{nn'}_\ms(\vec r_n, \vec r'_{n'}) = 2\pi \sum_{L,L'} \sgf^{nn'}_{LL'} R_L^n(r_n) R_{L'}^{n'}(r'_{n'}) Y_L(\uvec r_n) Y^*_{L'}(\uvec r'_{n'}) 
\end{equation}
Here, $R^n_L$ ($H^n_L$) is an (ir)regular partial wave solution to the radial Schr\"odinger equation at site $n$, with $L=(l,m)$ being the angular momentum numbers.
The coefficients $\sgf_{LL'}^{nn'}(z)$ are the matrix elements of the structural Green's function, which is the solution of a Dyson equation
\begin{equation}
    \sgf = \sgf_0 + \sgf_0 T \sgf
\end{equation}
where the $T$ is the single-site $t$-matrix with elements $\delta_{nn'}t^{nn}_{LL'}(z)$ and the matrix elements of $\sgf_0$ are the structure constants, which are evaluated using the real-space cluster method described in Ref.~\cite{LaraiaHEDP2021}.
Note, however, the use of different normalization conventions, as detailed in the Appendix.

The conductivity, Eq.~\eqref{eq:kg-kkr}, would be unnecessarily expensive to evaluate if the general expressions Eqs.~\eqref{eq:g-ss} and \eqref{eq:g-ms} were used directly.
To obtain a more practical expression for numerical implementation, we make the following simplifications.

First, we consider photoabsorption by core electrons only, defined to be those occupying deeply bound atomic-like orbitals.
This means we truncate the integral over initial-state energies in Eq.~\eqref{eq:kg-kkr} to only extend up to some $E_{\min}$ that lies below any valence bands, which we choose based on inspecting the density of states.
In this energy range, the Green's function for the initial state is replaced by core orbitals using Eq.~\eqref{eq:lehmann} and the Plemelj relation $\Im \frac{1}{E-E_a+i0^+} = -\pi\delta(E-E_a)$.
This collapses the energy integral and allows for the identification of independent contributions to the conductivity from each core electron.
As a further benefit, the double-sum over sites in Eq.~\eqref{eq:kg-kkr} need only include nuclear cells, since the potential in cells without nuclei is too shallow to support deeply bound electrons.

Second, we replace the dipole operators $\vec r_n + \vec R_n \to \vec r_n$, dropping the explicit dependence on the coordinates of the expansion centers.
In an isotropic plasma, this explicit coordinate dependence must vanish upon ensemble averaging anyway, so it is a waste of effort to retain it.
We also replace $\frac{1}{3}\vec r_n\cdot\vec r'_{n'} \to (\uvec z\cdot\vec r_n)(\vec r'_{n'}\cdot\uvec z)$, that is, we calculate the $zz$ Cartesian component of the conductivity tensor rather than one-third its trace. 
This is again permitted for isotropic systems like disordered plasmas, cutting the expense of the calculation by a factor of three.

Third, we adopt a muffin-tin approximation to the potential at each site.
This allows us drop the radial wave functions' dependence on the magnetic ($m$) quantum number and also allows the matrix elements of $T$ to be simplified to $\delta_{nn'}\delta_{LL'} t^n_l(z)$, where $t^n_l(z)$ is the familiar partial-wave $t$-matrix element for scattering by a spherically symmetric potential.

Fourth, the single-site term of the final-state Green's function is treated using a renormalization procedure suggested by Prange et al.~in Ref.~\cite{PrangePRB2009}.
This procedure separates the $r_n$ and $r'_n$ dependence of $G^{nn'}_\ss$, which are generally coupled together via $r_{\lessgtr}$.
This procedure is detailed and justified in the Appendix.
By separating the $r_n$ and $r'_n$ dependence of the Green's function, the spatial integrals in Eq.~\eqref{eq:kg-kkr} fully decouple into two radial and four angular integrals, and the angular integrals may be evaluated analytically.

With the above considerations, the optical conductivity for absorption by a core electron $a$ at site $n$ is given by
\begin{multline}
    \label{eq:sigma-a}
    \sigma_a^n(\omega) = \frac{2\pi e^2 \omega}{\Omega} \sum_{L,L'} [f(E^n_a) - f(E^n_a+\hbar\omega)] a_{LL'}^{nn}(E^n_a+\hbar\omega) \\ \times D^{n0}_{L_aL}(E_a^n,E_a^n+\hbar\omega) D^{n0}_{L'L_a}(E_a^n+\hbar\omega,E^n_a)
\end{multline}
where
\begin{equation}
\label{eq:a-coeff}
    a_{LL'}^{nn'}(z) = \delta_{nn'}\delta_{LL'} 
    - \frac{1}{2i} \left[\bar\sgf^{nn'}_{LL'}(z) - (-1)^{m+m'} \bar\sgf^{nn'*}_{\bar L\bar L'}(z)\right]
\end{equation}
are the spherical harmonic expansion coefficients of the imaginary part of the site-site Green's function
\begin{equation}
\label{eq:img-expand}
    \Im G^{nn'}(\vec r_n,\vec r'_{n'}) 
    = -2\pi\sum_{LL'}  a^{nn'}_{LL'} \bar R_l(r_n)  \bar R^{n'}_{l'}(r'_{n'}) Y_L(\uvec r_n) Y^*_{L'}(\uvec r'_{n'})
\end{equation}
In Eq.~\eqref{eq:a-coeff}, $\bar L = (l,-m)$ is the opposite-parity angular momentum index, and $\bar\sgf$ is a rescaled structural Green's function, defined in the Appendix.
The renormalized wave functions $\bar R_l^n$ are defined in the Appendix as well.
Also appearing in Eq.~\eqref{eq:sigma-a} is the dipole matrix element
\begin{equation}
    D^{nq}_{L_1L_2}(z_1,z_2)
    = \sqrt{\frac{4\pi}{3}} \int_{\Omega_n} r R^n_{l_1}(r;z_1) R^n_{l_2}(r;z_2) Y_{L_1}(\uvec r) Y^*_{L_2}(\uvec r) Y_{1q}(\uvec r) \dd{\vec r}
\end{equation}
which is evaluated by decomposing the cell $\Omega_n$ into the muffin-tin sphere and an interstitial region.
Over the muffin-tin sphere, the angular integrals may be performed analytically in terms of $3j$ symbols, and the usual dipole selection rules apply.
The remaining integral over the interstitial region is done using the quadrature rule described in Ref.\cite{AlamPRB2011}.
The interstitial region can often be neglected since core states decay rapidly away from the site center.

Equation~\eqref{eq:sigma-a} is appropriate for photoabsorption from a core state into a valence or continuum state, $E^n_a+\hbar\omega > E_{\min}$.
For transitions between core states $a\to b$, we may evaluate the simpler formula
\begin{equation}
    \label{eq:sigma-ab}
    \sigma_{ab}^n(\omega) = \frac{2\pi e^2}{\Omega} (E_b^n - E^n_a) (f_a - f_b) 
    |D^{n0}_{L_aL_b}(E^n_a,E^n_b)|^2 \delta(E^n_a - E^n_b + \hbar\omega)
\end{equation}
The Dirac delta is an artifact of the infinite lifetimes of the Kohn-Sham states and is in practice replaced by a unit Lorentzian line profile $\delta(E)\to (\Gamma/\pi)(E^2+\Gamma^2)^{-1}$, with the same width $\Gamma$ used to evaluate the retarded Green's function for core-valence transitions.
Since this width is a numerical parameter rather than a physical one, it is chosen to be much less than total line width due to the variation in transition energy from site to site.
  
The total conductivity due to core excitations at site $n$ is the sum of core-valence and core-core conductivities, for each core state that exists at site $n$.
The core-electron conductivity of a particular nuclear configuration is the sum from each site, which is then averaged over an ensemble of configurations produced from pseudoatom molecular dynamics (PAMD) simulations\cite{StarrettPRE2014}.
\comment{PAMD produces realistic nuclear configurations for warm and hot dense plasmas, and in this work we focus on conditions where it has validated against higher-fidelity approaches\cite{StarrettPRE2013,StarrettPRE2015}.}
Below, we give results in terms of the mass absorption coefficient
\begin{equation}
    \kappa(\omega) = \frac{\sigma(\omega)}{\rho c \epsilon_0}
\end{equation}
where $\rho$ is the mass density.
The plasma index of refraction has been assumed to be unity.
This is valid for all cases considered here, where the core-electron binding energy is on the $\si{\kilo\electronvolt}$ scale.
In comparison, the index of refraction is relevant only near or below the plasma frequency, $\omega_p=\sqrt{e^2n_e/\epsilon_0m_e}$, which is typically \SIrange{1}{100}{\electronvolt} for laser-produced plasmas.

\section{Results}
\label{sec:results}

\subsection{Comparison with an Average Atom Model}
\label{sec:vsaa}

The average atom (AA) model is a single-center approximation to the electronic structure of plasma ions. 
The Kohn-Sham equations are solved in the potential of a single nucleus, with free-particle boundary conditions imposed beyond the ion-sphere radius.
Compared to the present RSGF method, the AA model neglects multiple scattering, does not have site-to-site variation in the muffin-tin radii, and its Kohn-Sham potential does not account for the arrangement of nuclei.
As a result, AA models miss important qualitative effects of the plasma environment on the opacity.

\begin{figure}
    \centering
    \includegraphics[width=\columnwidth]{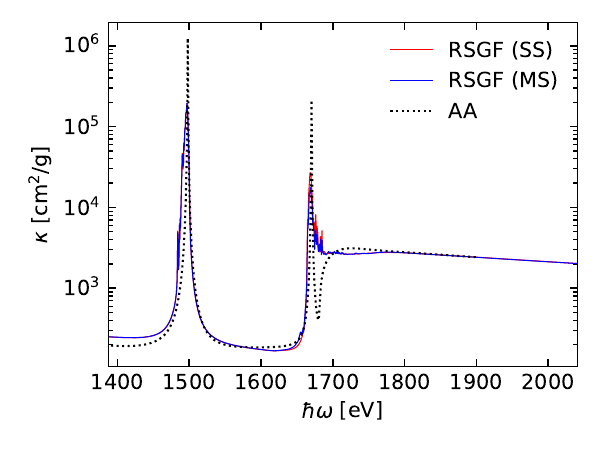}
    \caption{Opacity of aluminum at \SI{2.7}{\gram\per\cubic\cm} and \SI{100}{\electronvolt}. Solid curves are the results of RSGF in the single-site approximation (red) and including multiple scattering (blue). The black dotted curve is the result of the Tartarus average-atom model.}
    \label{fig:vsaa}
\end{figure}

This is illustrated in Fig.~\ref{fig:vsaa}, which compares the opacity of the Tartarus AA model\cite{StarrettCPC2019, GillPRE2021} to the RSGF method for an aluminum plasma at \SI{2.7}{\gram\per\cubic\centi\meter} and \SI{100}{\electronvolt}.
Both calculations use the KSDT local exchange-correlation functional\cite{KarasievPRL2014}. 
The RSGF calculations were averaged over 37 PAMD configurations, each containing 8 atoms. 
The photon energy range shown emphasizes the $K$-shell opacity but also includes the opacity from photoionization of the $L$-shell.
Two qualitative difference are worth noting.

First, the $1s$-$2p$ line is sharp in the AA model, but quite broad in the RSGF calculation.
In both approaches, the $1s$ and $2p$ are in the discrete spectrum (core states in the case of RSGF), and an artificial broadening of $\Gamma=\SI{0.27}{\electronvolt}$ has been applied.
The sharpness of the AA line comes from the fact that there is only one atom under consideration, and thus a single line.
In contrast, the RSGF calculation allows the $1s$ and $2p$ eigenvalues to vary site-to-site and across different arrangements of the nuclei. 
In doing so, the RSGF opacity accounts for the ion Stark effect, which is the dominant line broadening mechanisms in hot dense plasmas.

The second qualitative difference between the AA and RSGF comes near the $1s$ photoionization edge.
In the AA model, there is a well-defined edge as well as a well-defined $1s$-$3p$ line.
This occurs because the AA model has a hard threshold between its bound and continuous spectrum. 
Thus, the $3p$ orbital, despite being very weakly bound, still contributes a clear line in the AA opacity.
In contrast, the RSGF method treats the $3p$ electron as a valence state, which can lie either above or below the ionization threshold depending on each atom's local environment.
Thus, in RSGF, the $1s$ edge subsumes the $1s$-$3p$ line into single merged feature, a phenomenon which single-atom approaches can only capture by invoking external continuum lowering models.

Another interesting result is shown in Fig.~\ref{fig:vsaa}. 
We compare opacities from our full calculation (labeled RSGF (MS)) to those that set the multiple scattering term in the Green's function, Eq.~\eqref{eq:g-ms}, to zero (labeled RSGF SS).   
On the whole, these calculations agree very well, with only slight differences seen.
This means that for this case, the multiple scattering effect for opacity is washed out by the ion Stark effect, and can be safely ignored. 
However, this is not a general result and we would expect multiple scattering to be more important for strongly coupled ionic fluids (and solids), where site to site variation is smaller and the ion Stark effect less significant.

\subsection{Comparison with Plane-Wave DFT Calculations}
\label{sec:vspw}

Recently, Karasiev and Hu have published a systematic study of the opacity of silicon using plane-wave DFT\cite{KarasievPRE2021}.
They were able to perform calculations at both lower density and higher temperature than is usually feasible with plane-wave basis sets by supplementing multi-center calculations with single-atom calculations.
This makes for a valuable test of the RSGF method, since plane-wave and real-space DFT are formally equivalent but rely on very different approximations to be made practical.

\begin{figure}
    \centering
    \includegraphics[width=\columnwidth]{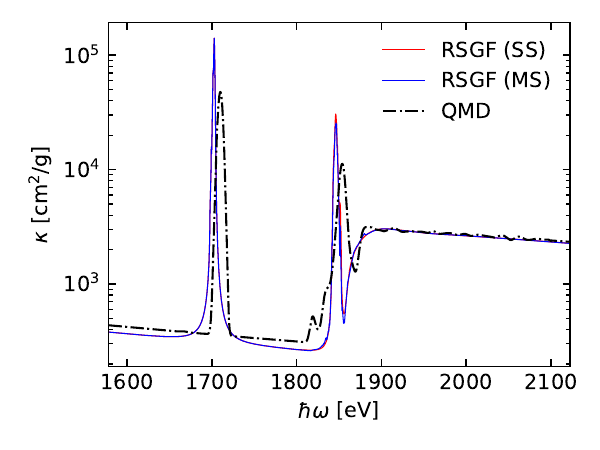}
    \caption{Opacity of silicon at \SI{1}{\gram\per\cubic\cm} and \SI{43.1}{\eV}. Solid curves are the results of RSGF in the single-site approximation (red) and including multiple scattering (blue). The black dash-dotted curve is the result of plane-wave DFT calculations by Karasiev and Hu\cite{KarasievPRE2021}.}
    \label{fig:vspw}
\end{figure}

We single out the case of silicon at \SI{1}{\gram\per\cubic\cm} and \SI{43.1}{\eV} (\SI{500}{\kilo\kelvin}).
At these conditions, both methods predict that silicon has a fully occupied $K$ shell, a partially occupied $L$ shell, and slightly occupied $M$ shell that is close to the continuum.
In Fig.~\ref{fig:vspw} we compare the opacity calculated in Ref.~\cite{KarasievPRE2021} with RSGF in both the single-site and multiple-scattering approximations.
The RSGF calculations made use of 18 snapshots from an 8-atom PAMD simulation plus 30 extra expansion centers.

Certain features of the spectrum are in good agreement between the methods.
The $1s\to2p$ lines near \SI{1700}{\eV} differ in transition energy by only \SIrange{10}{20}{\eV}.
This is easily within the variation one might expect from the use of different exchange-correlation functionals (local in the case of RSGF, versus gradient-corrected for the plane-wave results) and/or different treatments of the nuclear potential (Coulomb in RSGF versus PAW pseudopotential in plane-wave).
The $K$ edge is also in excellent agreement between the methods, with the only difference of note being the slightly steeper pre-edge predicted in the plane-wave calculation.

The most visible disagreement between the plane-wave and RSGF methods comes in the $K\to M$ lines predicted by the plane-wave calculation between \SIrange{1800}{1900}{\eV}.
The RSGF calculations both predict an unambiguous $1s\to3p$-like transition that is only slightly affected by multiple-scattering effects.
This would suggest that each atom of the RSGF calculation has a reasonably well-defined atomic-like $M$ shell, despite it being near the continuum.
In contrast, the plane-wave calculation predicts that the $M$ shell is significantly distorted by the plasma environment, signaled by the weak $1s\to3s$-like ``forbidden'' transition near \SI{1820}{\eV}.
The RSGF calculations do predict that such a transition occurs but with a much smaller cross-section, such that its contribution to the opacity is negligible.

The most likely cause of the observed difference is the use of a muffin-tin potential in the RSGF calculations combined with the insertion of non-nuclear spheres.
Together, these serve to make each cell nearly spherical.
In the muffin-tin approximation, spherical symmetry at each site can be broken in only two ways: either in the non-spherical interstitial region where the potential is constant or by accounting for multiple scattering.
The first effect is suppressed by the inclusion of non-nuclear spheres, which tends to shrink the size of the interstitial regions.
The second effect is suppressed by the fact that the structural Green's function is based on muffin-tin $t$-matrices, which are diagonal in angular momentum basis.
Thus, even though the solution of the Dyson equation couples together partial waves of different $l$ and $m$, this alone is not enough to adequately capture the deformation of the $M$ shell.
To capture these transitions within RSGF would require a so-called ``full-potential'' implementation \cite{asato1999full}, which does away with the muffin-tin construction and allows the single-site potentials to be non-spherical.

\subsection{Comparison with Pulse-Power Experiments}
\label{sec:vsz}

\begin{figure}
    \centering
    \includegraphics[height=0.8\textheight]{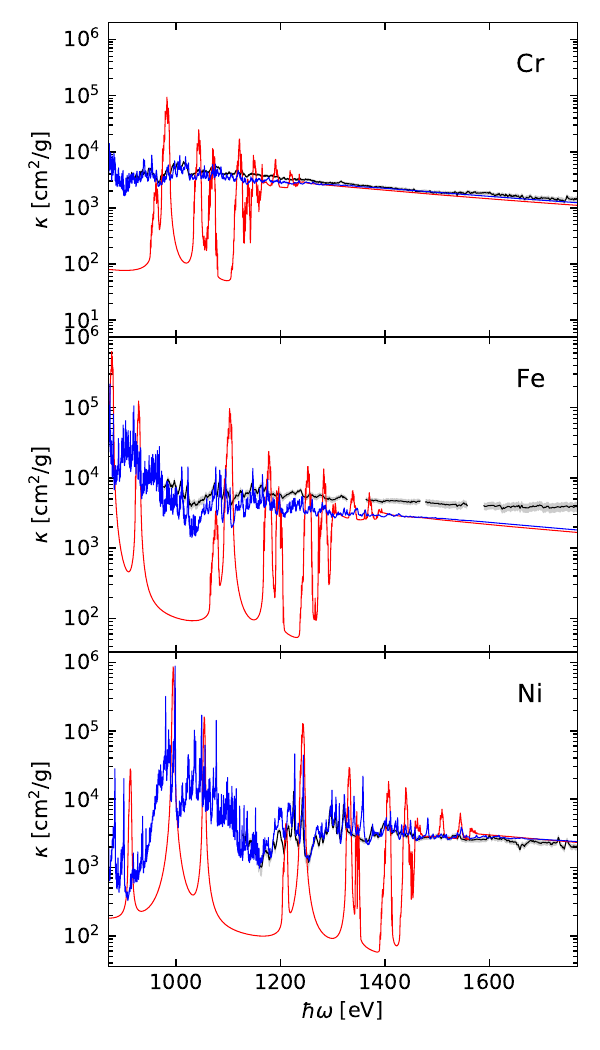}
    \caption{\comment{Opacity of chromium, iron, and nickel  \SIlist{0.16;0.16;0.17}{\gram\per\cubic\cm} respectively. Solid red curves are the results of RSGF at a temperature of \SI{180}{\eV} in the single-site approximation. The solid black curves show experimental results by Bailey et al.\cite{BaileyNL2015} and Nagayama et al.\cite{NagayamaPRL2019}. The solid blue curves are taken from the OPLIB database at a temperature \SI{175}{\eV}.} }
    \label{fig:vsz}
\end{figure}

Having demonstrated that the RSGF method is superior to average-atom calculations and in good agreement with plane-wave DFT calculations for hot dense plasmas, we now compare with the opacity experiments performed at Sandia National Laboratory on chromium, iron, and nickel\cite{BaileyNL2015,NagayamaPRL2019} \comment{as well as the OPLIB atomic opacity database\cite{ColganApJ2016}}.
For all three elements, the $L$-shell opacity was measured near conditions of \SI{180}{\eV} and \SIrange{0.16}{0.17}{\gram\per\cubic\cm}.
We have performed RSGF opacity calculations in the single-site approximation at these conditions based on 19 independent 8-atom molecular dynamics configurations.
To our knowledge, these are the first predictions based on multi-center electronic structure calculations.
Our results are shown in Fig.~\ref{fig:vsz}, alongside the experimental results.
There are two important observations to draw from this.

First, at high energies, where the opacity is dominated by $L$-shell photoionization, RSGF is in very good agreement with the experiments on chromium and nickel but severely underestimates the iron opacity.
This is consistent with the results \comment{from the Los Alamos OPLIB database, which is representative of the state of the art in the isolated-atom approach.}
We have also verified that accounting for multiple scattering in the RSGF calculations makes no discernible difference to the iron opacity prediction above \SI{1500}{\eV}. 
We conclude then that the high measured iron opacity compared to atomic models is not explained by taking better account of the the plasma environment.

The second important observation is the complete failure of the RSGF calculation to capture the measured bound-bound absorption lines.
Specifically, the RSGF calculations predict broad but strong bound-bound lines, with deep windows separating them, whereas experiments \comment{and isolated-atom calculations} indicate the bound-bound spectrum is made up of a dense sea of weaker lines.
This issue is not unique to RSGF; it is a fundamental problem with any opacity model based on finite-temperature DFT, including average-atom and plane-wave approaches.
The problem is that the eigenstates of finite-temperature DFT do not represent a single electron in an individual ion.
Rather, they represent a sort of mean eigenstate, taken over a thermal ensemble of ions of different charges and electronic configurations.
Each of these electronic configurations produces distinct absorption lines in reality, which are missed when one only considers transitions between the Kohn-Sham eigenstates.
This problem has long been recognized in the context of average-atom models, for which there exist several models for ``undoing'' the thermal averaging to predict the underlying statistical distribution of discrete-occupation ion configurations\cite{FaussurierPRE2018,SonPRX2014,PironHEDP2013,PerrotPA1988}.

\section{Conclusions}
\label{sec:conc}

In summary, the RSGF method has been shown to be an attractive means of predicting the x-ray spectra of warm and hot dense plasmas.
Being a multi-center approach, it naturally includes the effects of continuum lowering and Stark broadening, which are difficult to model reliably in single-atom opacity models.
Being a local real-space approach, it avoids many technical limitations of standard plane-wave DFT opacity calculations at high temperature related to pseuopotentials and the required basis set size.

Comparison with experimental opacity measurements for Cr and Ni reveals good agreement with the bound-free part of the spectrum, and strong disagreement with the bound-bound part.  This disagreement is an expected limitation of using the Kohn-Sham DFT eigenstates.  Comparison with measurements for Fe reveals that neither multiple-scattering or non-self-consistent plasma physics is the source of the difference between models and experiment in the bound-free part of the spectrum \cite{NagayamaPRL2019, krief2018effect}.

\appendix

\section{Normalization and Renormalization}

The site-site Green's function, Eq.~\eqref{eq:g-ss}, is built from solutions to the radial Schr\"odinger equation centered at site $n$, 
\begin{equation}
  \label{eq:schrod}
  \Bqty{\dv[2]{r} + \frac{2}{r} \dv{r} - \frac{l(l+1)}{r^2} - \frac{2m_e}{\hbar^2} V^n(r) + k^2 } \bmqty{R^n_l(r; z) \\ H^n_l(r;z)} = 0
\end{equation}
Here, $k = \sqrt{2m_ez/\hbar^2}$ is the complex momentum, and $V^n$ is the Kohn-Sham potential centered about $\vec R_n$ and treated in the muffin-tin approximation.
Since the complex energy wave functions are not square-integrable, they must be normalized as scattering states.
We adopt the convention that where the potential vanishes, the wave functions take the form 
\begin{equation}
  \label{eq:r-norm}
  R^n_l = \sqrt{\frac{2m_ek}{\pi\hbar}} \frac{1}{2}\bqty{e^{i\delta_l^n} h_l^+(kr_n) + e^{-i\delta_l^n} h_l^-(kr_n)}
\end{equation}
\begin{equation}
      H^n_l =\sqrt{\frac{2m_ek}{\pi\hbar}} e^{i\delta_l^n} h_l^+(kr_n)
\end{equation}
where $h^\pm_l$ are spherical Hankel functions.
The complex scattering phase shift $\delta_l^n(z)$ is determined by matching Eq.~\eqref{eq:r-norm} to the regular wave function obtained from integrating Eq.~\eqref{eq:schrod} outward to the muffin-tin radius.
The phase shift also determines the $T$-matrix element 
\begin{equation}
    t_l^n(z) = \frac{1-e^{2i\delta_l^n}}{2ik}
\end{equation}
used to solve for the structural Green's function.

Other normalizations of the wave functions are possible, but the choice made here offers a particular advantage for calculating optical properties, due to a procedure by Prange et al\cite{PrangePRB2009}, which makes the calculation of double-spatial integral in Eq.~\eqref{eq:kg-green} much simpler to evaluate.
This integral is cumbersome because of the product $R_l^n(r_<; z) H_l^n(r_>)$ appearing in the single-site Green's function.
However, on the positive real energy axis, the regular wave function is real and related to the irregular wave function by
\begin{equation}
  \Im[i H_l^n(r_n; E)] = R_l^n(r_n; E)
\end{equation}
Then when taking the imaginary part of the single-site Green's function, only regular wave functions are needed, and the spatial dependence is totally decoupled
\begin{equation}
  \Im G_\ss^{nn}(\vec r_n, \vec r'_n; E)
  = -2\pi \sum_L R_l^n(r_n;z) R_l^n(r'_n;z) Y_L(\uvec r_n) Y^*_L(\uvec r_n')
\end{equation}
This separation of the $r_n$ and $r'_n$ dependence greatly simplifies the calculation, but it does not hold away from the real axis. 
Instead, one supposes that there exist renormalized wave functions $\bar R_l^n$ that obey
\begin{equation}
    \Im[iR_l^n(r_<;z) H_l^n(r_>;z)] \approx \bar R_l^n(r_n;z) \bar R_l^n(r'_n;z)
\end{equation}
and which relate to the regular wave function by
\begin{equation}
    \bar R_l^n(r_m;z) = A^n_l(z) \Re R_l^n(r_n;z)
\end{equation}
The constant $A^n_l$ is chosen so that the renormalized wave functions preserve the single-site density of states, $\chi^n_\ss(z) \propto \int_{\Omega_n} \Im G_\ss^{nn}(r_n,r_n;z) \dd{\vec r_n}$, leading to
\begin{equation}
    \bqty{A^n_l(z)}^2 = \frac{\int_{\Omega_n} \Im[i R_l^n(r;z) H_l^n(r;z)] \dd{\vec r}}{\int_{\Omega_n} \bqty{\Re R_l^n(r;z)}^2 \dd{\vec r}}
\end{equation}

As for the multiple-scattering Green's function, its $\vec r_n$ and $\vec r'_{n'}$ dependence is already separated, so there is no need to introduce renormalized wave functions.
However for uniformity of notation in Eq.~\eqref{eq:img-expand}, it is helpful to use the renormalized wave functions anyway and to compensate for the extra $A$-factors by rescaling the structural Green's function
\begin{equation}
    \bar\sgf_{LL'}^{nn'} = \frac{e^{i(\delta_l^n + \delta_{l'}^{n'})}}{2kA_l^nA_{l'}^{n'}} \sgf_{LL'}^{nn'}
\end{equation}
The other factors compensate for the non-standard wave function normalization used here.

\acknowledgements

We wish to thank N.~M.~Gill for providing the average-atom data on Al, V.~V.~Karasiev for providing plane-wave DFT data on Si, and T.~Nagayama for providing the experimental data on Cr, Fe, and Ni. 

\bibliography{refs}

\begin{thebibliography}{52}%
\makeatletter
\providecommand \@ifxundefined [1]{%
 \@ifx{#1\undefined}
}%
\providecommand \@ifnum [1]{%
 \ifnum #1\expandafter \@firstoftwo
 \else \expandafter \@secondoftwo
 \fi
}%
\providecommand \@ifx [1]{%
 \ifx #1\expandafter \@firstoftwo
 \else \expandafter \@secondoftwo
 \fi
}%
\providecommand \natexlab [1]{#1}%
\providecommand \enquote  [1]{``#1''}%
\providecommand \bibnamefont  [1]{#1}%
\providecommand \bibfnamefont [1]{#1}%
\providecommand \citenamefont [1]{#1}%
\providecommand \href@noop [0]{\@secondoftwo}%
\providecommand \href [0]{\begingroup \@sanitize@url \@href}%
\providecommand \@href[1]{\@@startlink{#1}\@@href}%
\providecommand \@@href[1]{\endgroup#1\@@endlink}%
\providecommand \@sanitize@url [0]{\catcode `\\12\catcode `\$12\catcode
  `\&12\catcode `\#12\catcode `\^12\catcode `\_12\catcode `\%12\relax}%
\providecommand \@@startlink[1]{}%
\providecommand \@@endlink[0]{}%
\providecommand \url  [0]{\begingroup\@sanitize@url \@url }%
\providecommand \@url [1]{\endgroup\@href {#1}{\urlprefix }}%
\providecommand \urlprefix  [0]{URL }%
\providecommand \Eprint [0]{\href }%
\providecommand \doibase [0]{http://dx.doi.org/}%
\providecommand \selectlanguage [0]{\@gobble}%
\providecommand \bibinfo  [0]{\@secondoftwo}%
\providecommand \bibfield  [0]{\@secondoftwo}%
\providecommand \translation [1]{[#1]}%
\providecommand \BibitemOpen [0]{}%
\providecommand \bibitemStop [0]{}%
\providecommand \bibitemNoStop [0]{.\EOS\space}%
\providecommand \EOS [0]{\spacefactor3000\relax}%
\providecommand \BibitemShut  [1]{\csname bibitem#1\endcsname}%
\let\auto@bib@innerbib\@empty
\bibitem [{\citenamefont {Fontes}\ \emph {et~al.}(2015)\citenamefont {Fontes},
  \citenamefont {Zhang}, \citenamefont {Abdallah~Jr.}, \citenamefont {Clark},
  \citenamefont {Kilcrease}, \citenamefont {Colgan}, \citenamefont
  {Cunningham}, \citenamefont {Hakel}, \citenamefont {Magee},\ and\
  \citenamefont {Sherrill}}]{FontesJPB2015}%
  \BibitemOpen
  \bibfield  {author} {\bibinfo {author} {\bibfnamefont {C.~J.}\ \bibnamefont
  {Fontes}}, \bibinfo {author} {\bibfnamefont {H.~L.}\ \bibnamefont {Zhang}},
  \bibinfo {author} {\bibfnamefont {J.}~\bibnamefont {Abdallah~Jr.}}, \bibinfo
  {author} {\bibfnamefont {R.~E.~H.}\ \bibnamefont {Clark}}, \bibinfo {author}
  {\bibfnamefont {D.~P.}\ \bibnamefont {Kilcrease}}, \bibinfo {author}
  {\bibfnamefont {J.}~\bibnamefont {Colgan}}, \bibinfo {author} {\bibfnamefont
  {R.~T.}\ \bibnamefont {Cunningham}}, \bibinfo {author} {\bibfnamefont
  {P.}~\bibnamefont {Hakel}}, \bibinfo {author} {\bibfnamefont {N.~H.}\
  \bibnamefont {Magee}}, \ and\ \bibinfo {author} {\bibfnamefont {M.~E.}\
  \bibnamefont {Sherrill}},\ }\href@noop {} {\bibfield  {journal} {\bibinfo
  {journal} {Journal of Physics B: Atomic, Molecular and Optical Physics}\
  }\textbf {\bibinfo {volume} {48}},\ \bibinfo {pages} {144014} (\bibinfo
  {year} {2015})}\BibitemShut {NoStop}%
\bibitem [{\citenamefont {Iglesias}\ and\ \citenamefont
  {Rogers}(1996)}]{IglesiasAJ1996}%
  \BibitemOpen
  \bibfield  {author} {\bibinfo {author} {\bibfnamefont {C.~A.}\ \bibnamefont
  {Iglesias}}\ and\ \bibinfo {author} {\bibfnamefont {F.~J.}\ \bibnamefont
  {Rogers}},\ }\href@noop {} {\bibfield  {journal} {\bibinfo  {journal} {The
  astrophysical journal}\ }\textbf {\bibinfo {volume} {464}},\ \bibinfo {pages}
  {943} (\bibinfo {year} {1996})}\BibitemShut {NoStop}%
\bibitem [{\citenamefont {Badnell}\ \emph {et~al.}(2005)\citenamefont
  {Badnell}, \citenamefont {Bautista}, \citenamefont {Butler}, \citenamefont
  {Delahaye}, \citenamefont {Mendoza}, \citenamefont {Palmeri}, \citenamefont
  {Zeippen},\ and\ \citenamefont {Seaton}}]{badnellMNRAS2005}%
  \BibitemOpen
  \bibfield  {author} {\bibinfo {author} {\bibfnamefont {N.~R.}\ \bibnamefont
  {Badnell}}, \bibinfo {author} {\bibfnamefont {M.~A.}\ \bibnamefont
  {Bautista}}, \bibinfo {author} {\bibfnamefont {K.}~\bibnamefont {Butler}},
  \bibinfo {author} {\bibfnamefont {F.}~\bibnamefont {Delahaye}}, \bibinfo
  {author} {\bibfnamefont {C.}~\bibnamefont {Mendoza}}, \bibinfo {author}
  {\bibfnamefont {P.}~\bibnamefont {Palmeri}}, \bibinfo {author} {\bibfnamefont
  {C.~J.}\ \bibnamefont {Zeippen}}, \ and\ \bibinfo {author} {\bibfnamefont
  {M.~J.}\ \bibnamefont {Seaton}},\ }\href@noop {} {\bibfield  {journal}
  {\bibinfo  {journal} {Monthly Notices of the Royal Astronomical Society}\
  }\textbf {\bibinfo {volume} {360}},\ \bibinfo {pages} {458} (\bibinfo {year}
  {2005})}\BibitemShut {NoStop}%
\bibitem [{\citenamefont {Aberg}\ \emph {et~al.}(2020)\citenamefont {Aberg},
  \citenamefont {Fenn}, \citenamefont {Foord}, \citenamefont {Grabowski},
  \citenamefont {Iglesias},\ and\ \citenamefont {Wilson}}]{OpusReport}%
  \BibitemOpen
  \bibfield  {author} {\bibinfo {author} {\bibfnamefont {P.~D.}\ \bibnamefont
  {Aberg}}, \bibinfo {author} {\bibfnamefont {D.}~\bibnamefont {Fenn}},
  \bibinfo {author} {\bibfnamefont {M.~E.}\ \bibnamefont {Foord}}, \bibinfo
  {author} {\bibfnamefont {P.~E.}\ \bibnamefont {Grabowski}}, \bibinfo {author}
  {\bibfnamefont {C.~A.}\ \bibnamefont {Iglesias}}, \ and\ \bibinfo {author}
  {\bibfnamefont {B.~G.}\ \bibnamefont {Wilson}},\ }\href {\doibase
  10.2172/1650424} {\  (\bibinfo {year} {2020}),\ 10.2172/1650424}\BibitemShut
  {NoStop}%
\bibitem [{\citenamefont {Gill}\ \emph {et~al.}(2021)\citenamefont {Gill},
  \citenamefont {Fontes},\ and\ \citenamefont {Starrett}}]{GillPRE2021}%
  \BibitemOpen
  \bibfield  {author} {\bibinfo {author} {\bibfnamefont {N.~M.}\ \bibnamefont
  {Gill}}, \bibinfo {author} {\bibfnamefont {C.~J.}\ \bibnamefont {Fontes}}, \
  and\ \bibinfo {author} {\bibfnamefont {C.~E.}\ \bibnamefont {Starrett}},\
  }\href {\doibase 10.1103/PhysRevE.103.043206} {\bibfield  {journal} {\bibinfo
   {journal} {Phys. Rev. E}\ }\textbf {\bibinfo {volume} {103}},\ \bibinfo
  {pages} {043206} (\bibinfo {year} {2021})}\BibitemShut {NoStop}%
\bibitem [{\citenamefont {Ciricosta}\ \emph {et~al.}(2012)\citenamefont
  {Ciricosta}, \citenamefont {Vinko}, \citenamefont {Chung}, \citenamefont
  {Cho}, \citenamefont {Brown}, \citenamefont {Burian}, \citenamefont
  {Chalupsk{\`y}}, \citenamefont {Engelhorn}, \citenamefont {Falcone},
  \citenamefont {Graves} \emph {et~al.}}]{ciricostaPRL2012}%
  \BibitemOpen
  \bibfield  {author} {\bibinfo {author} {\bibfnamefont {O.}~\bibnamefont
  {Ciricosta}}, \bibinfo {author} {\bibfnamefont {S.~M.}\ \bibnamefont
  {Vinko}}, \bibinfo {author} {\bibfnamefont {H.-K.}\ \bibnamefont {Chung}},
  \bibinfo {author} {\bibfnamefont {B.-I.}\ \bibnamefont {Cho}}, \bibinfo
  {author} {\bibfnamefont {C.~R.~D.}\ \bibnamefont {Brown}}, \bibinfo {author}
  {\bibfnamefont {T.}~\bibnamefont {Burian}}, \bibinfo {author} {\bibfnamefont
  {J.}~\bibnamefont {Chalupsk{\`y}}}, \bibinfo {author} {\bibfnamefont
  {K.}~\bibnamefont {Engelhorn}}, \bibinfo {author} {\bibfnamefont {R.~W.}\
  \bibnamefont {Falcone}}, \bibinfo {author} {\bibfnamefont {C.}~\bibnamefont
  {Graves}},  \emph {et~al.},\ }\href@noop {} {\bibfield  {journal} {\bibinfo
  {journal} {Physical review letters}\ }\textbf {\bibinfo {volume} {109}},\
  \bibinfo {pages} {065002} (\bibinfo {year} {2012})}\BibitemShut {NoStop}%
\bibitem [{\citenamefont {Iglesias}\ and\ \citenamefont
  {Sterne}(2013)}]{iglesias2013fluctuations}%
  \BibitemOpen
  \bibfield  {author} {\bibinfo {author} {\bibfnamefont {C.~A.}\ \bibnamefont
  {Iglesias}}\ and\ \bibinfo {author} {\bibfnamefont {P.~A.}\ \bibnamefont
  {Sterne}},\ }\href@noop {} {\bibfield  {journal} {\bibinfo  {journal} {High
  Energy Density Physics}\ }\textbf {\bibinfo {volume} {9}},\ \bibinfo {pages}
  {103} (\bibinfo {year} {2013})}\BibitemShut {NoStop}%
\bibitem [{\citenamefont {Bailey}\ \emph {et~al.}(2015)\citenamefont {Bailey},
  \citenamefont {Nagayama}, \citenamefont {Loisel}, \citenamefont {Rochau},
  \citenamefont {Blancard}, \citenamefont {Colgan}, \citenamefont {Cosse},
  \citenamefont {Faussurier}, \citenamefont {Fontes}, \citenamefont {Gilleron},
  \citenamefont {Golovkin}, \citenamefont {Hansen}, \citenamefont {Iglesias},
  \citenamefont {Kilcrease}, \citenamefont {MacFarlane}, \citenamefont
  {Mancini}, \citenamefont {Nahar}, \citenamefont {Orban}, \citenamefont
  {Pain}, \citenamefont {Pradhan}, \citenamefont {Sherrill},\ and\
  \citenamefont {Wilson}}]{BaileyNL2015}%
  \BibitemOpen
  \bibfield  {author} {\bibinfo {author} {\bibfnamefont {J.~E.}\ \bibnamefont
  {Bailey}}, \bibinfo {author} {\bibfnamefont {T.}~\bibnamefont {Nagayama}},
  \bibinfo {author} {\bibfnamefont {G.~P.}\ \bibnamefont {Loisel}}, \bibinfo
  {author} {\bibfnamefont {G.~A.}\ \bibnamefont {Rochau}}, \bibinfo {author}
  {\bibfnamefont {C.}~\bibnamefont {Blancard}}, \bibinfo {author}
  {\bibfnamefont {J.}~\bibnamefont {Colgan}}, \bibinfo {author} {\bibfnamefont
  {P.}~\bibnamefont {Cosse}}, \bibinfo {author} {\bibfnamefont
  {G.}~\bibnamefont {Faussurier}}, \bibinfo {author} {\bibfnamefont {C.~J.}\
  \bibnamefont {Fontes}}, \bibinfo {author} {\bibfnamefont {F.}~\bibnamefont
  {Gilleron}}, \bibinfo {author} {\bibfnamefont {I.}~\bibnamefont {Golovkin}},
  \bibinfo {author} {\bibfnamefont {S.~B.}\ \bibnamefont {Hansen}}, \bibinfo
  {author} {\bibfnamefont {C.~A.}\ \bibnamefont {Iglesias}}, \bibinfo {author}
  {\bibfnamefont {D.~P.}\ \bibnamefont {Kilcrease}}, \bibinfo {author}
  {\bibfnamefont {J.~J.}\ \bibnamefont {MacFarlane}}, \bibinfo {author}
  {\bibfnamefont {R.~C.}\ \bibnamefont {Mancini}}, \bibinfo {author}
  {\bibfnamefont {S.~N.}\ \bibnamefont {Nahar}}, \bibinfo {author}
  {\bibfnamefont {C.}~\bibnamefont {Orban}}, \bibinfo {author} {\bibfnamefont
  {J.-C.}\ \bibnamefont {Pain}}, \bibinfo {author} {\bibfnamefont {A.~K.}\
  \bibnamefont {Pradhan}}, \bibinfo {author} {\bibfnamefont {M.}~\bibnamefont
  {Sherrill}}, \ and\ \bibinfo {author} {\bibfnamefont {B.~G.}\ \bibnamefont
  {Wilson}},\ }\href@noop {} {\bibfield  {journal} {\bibinfo  {journal}
  {Nature}\ }\textbf {\bibinfo {volume} {517}},\ \bibinfo {pages} {56}
  (\bibinfo {year} {2015})}\BibitemShut {NoStop}%
\bibitem [{\citenamefont {Hu}\ \emph {et~al.}(2017)\citenamefont {Hu} \emph
  {et~al.}}]{huPRL2017}%
  \BibitemOpen
  \bibfield  {author} {\bibinfo {author} {\bibfnamefont {S.~X.}\ \bibnamefont
  {Hu}} \emph {et~al.},\ }\href@noop {} {\bibfield  {journal} {\bibinfo
  {journal} {Physical review letters}\ }\textbf {\bibinfo {volume} {119}},\
  \bibinfo {pages} {065001} (\bibinfo {year} {2017})}\BibitemShut {NoStop}%
\bibitem [{\citenamefont {Massacrier}\ \emph {et~al.}(2021)\citenamefont
  {Massacrier}, \citenamefont {B\"ohme}, \citenamefont {Vorberger},
  \citenamefont {Soubiran},\ and\ \citenamefont
  {Militzer}}]{massacrierPRR2021}%
  \BibitemOpen
  \bibfield  {author} {\bibinfo {author} {\bibfnamefont {G.}~\bibnamefont
  {Massacrier}}, \bibinfo {author} {\bibfnamefont {M.}~\bibnamefont {B\"ohme}},
  \bibinfo {author} {\bibfnamefont {J.}~\bibnamefont {Vorberger}}, \bibinfo
  {author} {\bibfnamefont {F.}~\bibnamefont {Soubiran}}, \ and\ \bibinfo
  {author} {\bibfnamefont {B.}~\bibnamefont {Militzer}},\ }\href {\doibase
  10.1103/PhysRevResearch.3.023026} {\bibfield  {journal} {\bibinfo  {journal}
  {Phys. Rev. Research}\ }\textbf {\bibinfo {volume} {3}},\ \bibinfo {pages}
  {023026} (\bibinfo {year} {2021})}\BibitemShut {NoStop}%
\bibitem [{\citenamefont {Son}\ \emph {et~al.}(2014)\citenamefont {Son},
  \citenamefont {Thiele}, \citenamefont {Jurek}, \citenamefont {Ziaja},\ and\
  \citenamefont {Santra}}]{SonPRX1014}%
  \BibitemOpen
  \bibfield  {author} {\bibinfo {author} {\bibfnamefont {S.-K.}\ \bibnamefont
  {Son}}, \bibinfo {author} {\bibfnamefont {R.}~\bibnamefont {Thiele}},
  \bibinfo {author} {\bibfnamefont {Z.}~\bibnamefont {Jurek}}, \bibinfo
  {author} {\bibfnamefont {B.}~\bibnamefont {Ziaja}}, \ and\ \bibinfo {author}
  {\bibfnamefont {R.}~\bibnamefont {Santra}},\ }\href {\doibase
  10.1103/PhysRevX.4.031004} {\bibfield  {journal} {\bibinfo  {journal} {Phys.
  Rev. X}\ }\textbf {\bibinfo {volume} {4}},\ \bibinfo {pages} {031004}
  (\bibinfo {year} {2014})}\BibitemShut {NoStop}%
\bibitem [{\citenamefont {Hohenberg}\ and\ \citenamefont
  {Kohn}(1964)}]{hohenberg1964inhomogeneous}%
  \BibitemOpen
  \bibfield  {author} {\bibinfo {author} {\bibfnamefont {P.}~\bibnamefont
  {Hohenberg}}\ and\ \bibinfo {author} {\bibfnamefont {W.}~\bibnamefont
  {Kohn}},\ }\href@noop {} {\bibfield  {journal} {\bibinfo  {journal} {Physical
  review}\ }\textbf {\bibinfo {volume} {136}},\ \bibinfo {pages} {B864}
  (\bibinfo {year} {1964})}\BibitemShut {NoStop}%
\bibitem [{\citenamefont {Kohn}\ and\ \citenamefont {Sham}(1965)}]{kohn65self}%
  \BibitemOpen
  \bibfield  {author} {\bibinfo {author} {\bibfnamefont {W.}~\bibnamefont
  {Kohn}}\ and\ \bibinfo {author} {\bibfnamefont {L.~J.}\ \bibnamefont
  {Sham}},\ }\href {\doibase 10.1103/PhysRev.140.A1133} {\bibfield  {journal}
  {\bibinfo  {journal} {Phys. Rev.}\ }\textbf {\bibinfo {volume} {140}},\
  \bibinfo {pages} {A1133} (\bibinfo {year} {1965})}\BibitemShut {NoStop}%
\bibitem [{\citenamefont {Mermin}(1965)}]{merminPR1965}%
  \BibitemOpen
  \bibfield  {author} {\bibinfo {author} {\bibfnamefont {N.~D.}\ \bibnamefont
  {Mermin}},\ }\href {\doibase 10.1103/PhysRev.137.A1441} {\bibfield  {journal}
  {\bibinfo  {journal} {Phys. Rev.}\ }\textbf {\bibinfo {volume} {137}},\
  \bibinfo {pages} {A1441} (\bibinfo {year} {1965})}\BibitemShut {NoStop}%
\bibitem [{\citenamefont {Rehr}\ and\ \citenamefont
  {Albers}(2000)}]{rehrRMP2000}%
  \BibitemOpen
  \bibfield  {author} {\bibinfo {author} {\bibfnamefont {J.~J.}\ \bibnamefont
  {Rehr}}\ and\ \bibinfo {author} {\bibfnamefont {R.~C.}\ \bibnamefont
  {Albers}},\ }\href {\doibase 10.1103/RevModPhys.72.621} {\bibfield  {journal}
  {\bibinfo  {journal} {Rev. Mod. Phys.}\ }\textbf {\bibinfo {volume} {72}},\
  \bibinfo {pages} {621} (\bibinfo {year} {2000})}\BibitemShut {NoStop}%
\bibitem [{\citenamefont {Mattern}\ \emph {et~al.}(2012)\citenamefont
  {Mattern}, \citenamefont {Seidler}, \citenamefont {Kas}, \citenamefont
  {Pacold},\ and\ \citenamefont {Rehr}}]{mattern2012real}%
  \BibitemOpen
  \bibfield  {author} {\bibinfo {author} {\bibfnamefont {B.~A.}\ \bibnamefont
  {Mattern}}, \bibinfo {author} {\bibfnamefont {G.~T.}\ \bibnamefont
  {Seidler}}, \bibinfo {author} {\bibfnamefont {J.~J.}\ \bibnamefont {Kas}},
  \bibinfo {author} {\bibfnamefont {J.~I.}\ \bibnamefont {Pacold}}, \ and\
  \bibinfo {author} {\bibfnamefont {J.~J.}\ \bibnamefont {Rehr}},\ }\href@noop
  {} {\bibfield  {journal} {\bibinfo  {journal} {Physical Review B}\ }\textbf
  {\bibinfo {volume} {85}},\ \bibinfo {pages} {115135} (\bibinfo {year}
  {2012})}\BibitemShut {NoStop}%
\bibitem [{\citenamefont {Mazevet}\ \emph {et~al.}(2005)\citenamefont
  {Mazevet}, \citenamefont {Desjarlais}, \citenamefont {Collins}, \citenamefont
  {Kress},\ and\ \citenamefont {Magee}}]{mazevet2005simulations}%
  \BibitemOpen
  \bibfield  {author} {\bibinfo {author} {\bibfnamefont {S.}~\bibnamefont
  {Mazevet}}, \bibinfo {author} {\bibfnamefont {M.~P.}\ \bibnamefont
  {Desjarlais}}, \bibinfo {author} {\bibfnamefont {L.~A.}\ \bibnamefont
  {Collins}}, \bibinfo {author} {\bibfnamefont {J.~D.}\ \bibnamefont {Kress}},
  \ and\ \bibinfo {author} {\bibfnamefont {N.~H.}\ \bibnamefont {Magee}},\
  }\href@noop {} {\bibfield  {journal} {\bibinfo  {journal} {Physical Review
  E}\ }\textbf {\bibinfo {volume} {71}},\ \bibinfo {pages} {016409} (\bibinfo
  {year} {2005})}\BibitemShut {NoStop}%
\bibitem [{\citenamefont {{Recoules}}\ and\ \citenamefont
  {{Mazevet}}(2009)}]{recoulesPRB2009}%
  \BibitemOpen
  \bibfield  {author} {\bibinfo {author} {\bibfnamefont {V.}~\bibnamefont
  {{Recoules}}}\ and\ \bibinfo {author} {\bibfnamefont {S.}~\bibnamefont
  {{Mazevet}}},\ }\href {\doibase 10.1103/PhysRevB.80.064110} {\bibfield
  {journal} {\bibinfo  {journal} {\prb}\ }\textbf {\bibinfo {volume} {80}},\
  \bibinfo {eid} {064110} (\bibinfo {year} {2009})}\BibitemShut {NoStop}%
\bibitem [{\citenamefont {{Desjarlais}}\ \emph {et~al.}(2002)\citenamefont
  {{Desjarlais}}, \citenamefont {{Kress}},\ and\ \citenamefont
  {{Collins}}}]{DesjarlaisPRE2002}%
  \BibitemOpen
  \bibfield  {author} {\bibinfo {author} {\bibfnamefont {M.~P.}\ \bibnamefont
  {{Desjarlais}}}, \bibinfo {author} {\bibfnamefont {J.~D.}\ \bibnamefont
  {{Kress}}}, \ and\ \bibinfo {author} {\bibfnamefont {L.~A.}\ \bibnamefont
  {{Collins}}},\ }\href {\doibase 10.1103/PhysRevE.66.025401} {\bibfield
  {journal} {\bibinfo  {journal} {\pre}\ }\textbf {\bibinfo {volume} {66}},\
  \bibinfo {eid} {025401} (\bibinfo {year} {2002})}\BibitemShut {NoStop}%
\bibitem [{\citenamefont {Starrett}\ and\ \citenamefont
  {Shaffer}(2020)}]{StarrettPRE2020}%
  \BibitemOpen
  \bibfield  {author} {\bibinfo {author} {\bibfnamefont {C.~E.}\ \bibnamefont
  {Starrett}}\ and\ \bibinfo {author} {\bibfnamefont {N.}~\bibnamefont
  {Shaffer}},\ }\href {\doibase 10.1103/PhysRevE.102.043211} {\bibfield
  {journal} {\bibinfo  {journal} {Phys. Rev. E}\ }\textbf {\bibinfo {volume}
  {102}},\ \bibinfo {pages} {043211} (\bibinfo {year} {2020})}\BibitemShut
  {NoStop}%
\bibitem [{\citenamefont {{Laraia}}\ \emph {et~al.}(2021)\citenamefont
  {{Laraia}}, \citenamefont {{Hansen}}, \citenamefont {{Shaffer}},
  \citenamefont {{Saumon}}, \citenamefont {{Kilcrease}},\ and\ \citenamefont
  {{Starrett}}}]{LaraiaHEDP2021}%
  \BibitemOpen
  \bibfield  {author} {\bibinfo {author} {\bibfnamefont {M.}~\bibnamefont
  {{Laraia}}}, \bibinfo {author} {\bibfnamefont {C.}~\bibnamefont {{Hansen}}},
  \bibinfo {author} {\bibfnamefont {N.~R.}\ \bibnamefont {{Shaffer}}}, \bibinfo
  {author} {\bibfnamefont {D.}~\bibnamefont {{Saumon}}}, \bibinfo {author}
  {\bibfnamefont {D.~P.}\ \bibnamefont {{Kilcrease}}}, \ and\ \bibinfo {author}
  {\bibfnamefont {C.~E.}\ \bibnamefont {{Starrett}}},\ }\href {\doibase
  10.1016/j.hedp.2021.100940} {\bibfield  {journal} {\bibinfo  {journal} {High
  Energy Density Physics}\ }\textbf {\bibinfo {volume} {40}},\ \bibinfo {eid}
  {100940} (\bibinfo {year} {2021})},\ \Eprint
  {http://arxiv.org/abs/2101.02088} {arXiv:2101.02088 [physics.plasm-ph]}
  \BibitemShut {NoStop}%
\bibitem [{\citenamefont {Ebert}\ \emph {et~al.}(2011)\citenamefont {Ebert},
  \citenamefont {Koedderitzsch},\ and\ \citenamefont {Minar}}]{EbertRPP2011}%
  \BibitemOpen
  \bibfield  {author} {\bibinfo {author} {\bibfnamefont {H.}~\bibnamefont
  {Ebert}}, \bibinfo {author} {\bibfnamefont {D.}~\bibnamefont
  {Koedderitzsch}}, \ and\ \bibinfo {author} {\bibfnamefont {J.}~\bibnamefont
  {Minar}},\ }\href@noop {} {\bibfield  {journal} {\bibinfo  {journal} {Reports
  on Progress in Physics}\ }\textbf {\bibinfo {volume} {74}},\ \bibinfo {pages}
  {096501} (\bibinfo {year} {2011})}\BibitemShut {NoStop}%
\bibitem [{\citenamefont {{Faulkner}}\ \emph {et~al.}(2018)\citenamefont
  {{Faulkner}}, \citenamefont {{Stocks}},\ and\ \citenamefont
  {{Wang}}}]{faulknerBook2018}%
  \BibitemOpen
  \bibfield  {author} {\bibinfo {author} {\bibfnamefont {J.~S.}\ \bibnamefont
  {{Faulkner}}}, \bibinfo {author} {\bibfnamefont {G.~M.}\ \bibnamefont
  {{Stocks}}}, \ and\ \bibinfo {author} {\bibfnamefont {Y.}~\bibnamefont
  {{Wang}}},\ }\href {\doibase 10.1088/2053-2563/aae7d8} {\emph {\bibinfo
  {title} {{Multiple Scattering Theory; Electronic structure of solids}}}}\
  (\bibinfo {year} {2018})\BibitemShut {NoStop}%
\bibitem [{\citenamefont {Korringa}(1947)}]{KorringaP1947}%
  \BibitemOpen
  \bibfield  {author} {\bibinfo {author} {\bibfnamefont {J.}~\bibnamefont
  {Korringa}},\ }\href@noop {} {\bibfield  {journal} {\bibinfo  {journal}
  {Physica}\ }\textbf {\bibinfo {volume} {13}},\ \bibinfo {pages} {392}
  (\bibinfo {year} {1947})}\BibitemShut {NoStop}%
\bibitem [{\citenamefont {Kohn}\ and\ \citenamefont
  {Rostoker}(1954)}]{KohnPR1954}%
  \BibitemOpen
  \bibfield  {author} {\bibinfo {author} {\bibfnamefont {W.}~\bibnamefont
  {Kohn}}\ and\ \bibinfo {author} {\bibfnamefont {N.}~\bibnamefont
  {Rostoker}},\ }\href@noop {} {\bibfield  {journal} {\bibinfo  {journal}
  {Physical Review}\ }\textbf {\bibinfo {volume} {94}},\ \bibinfo {pages}
  {1111} (\bibinfo {year} {1954})}\BibitemShut {NoStop}%
\bibitem [{\citenamefont {Wang}\ \emph {et~al.}(1995)\citenamefont {Wang},
  \citenamefont {Stocks}, \citenamefont {Shelton}, \citenamefont {Nicholson},
  \citenamefont {Szotek},\ and\ \citenamefont {Temmerman}}]{wangRPL1995}%
  \BibitemOpen
  \bibfield  {author} {\bibinfo {author} {\bibfnamefont {Y.}~\bibnamefont
  {Wang}}, \bibinfo {author} {\bibfnamefont {G.~M.}\ \bibnamefont {Stocks}},
  \bibinfo {author} {\bibfnamefont {W.~A.}\ \bibnamefont {Shelton}}, \bibinfo
  {author} {\bibfnamefont {D.~M.~C.}\ \bibnamefont {Nicholson}}, \bibinfo
  {author} {\bibfnamefont {Z.}~\bibnamefont {Szotek}}, \ and\ \bibinfo {author}
  {\bibfnamefont {W.~M.}\ \bibnamefont {Temmerman}},\ }\href {\doibase
  10.1103/PhysRevLett.75.2867} {\bibfield  {journal} {\bibinfo  {journal}
  {Phys. Rev. Lett.}\ }\textbf {\bibinfo {volume} {75}},\ \bibinfo {pages}
  {2867} (\bibinfo {year} {1995})}\BibitemShut {NoStop}%
\bibitem [{\citenamefont {Korringa}(1994)}]{KorringaPR1994}%
  \BibitemOpen
  \bibfield  {author} {\bibinfo {author} {\bibfnamefont {J.}~\bibnamefont
  {Korringa}},\ }\href {\doibase https://doi.org/10.1016/0370-1573(94)90122-8}
  {\bibfield  {journal} {\bibinfo  {journal} {Physics Reports}\ }\textbf
  {\bibinfo {volume} {238}},\ \bibinfo {pages} {341} (\bibinfo {year}
  {1994})}\BibitemShut {NoStop}%
\bibitem [{\citenamefont {Ankudinov}\ \emph {et~al.}(1998)\citenamefont
  {Ankudinov}, \citenamefont {Ravel}, \citenamefont {Rehr},\ and\ \citenamefont
  {Conradson}}]{ankudinovPRB1998}%
  \BibitemOpen
  \bibfield  {author} {\bibinfo {author} {\bibfnamefont {A.~L.}\ \bibnamefont
  {Ankudinov}}, \bibinfo {author} {\bibfnamefont {B.}~\bibnamefont {Ravel}},
  \bibinfo {author} {\bibfnamefont {J.~J.}\ \bibnamefont {Rehr}}, \ and\
  \bibinfo {author} {\bibfnamefont {S.~D.}\ \bibnamefont {Conradson}},\ }\href
  {\doibase 10.1103/PhysRevB.58.7565} {\bibfield  {journal} {\bibinfo
  {journal} {Phys. Rev. B}\ }\textbf {\bibinfo {volume} {58}},\ \bibinfo
  {pages} {7565} (\bibinfo {year} {1998})}\BibitemShut {NoStop}%
\bibitem [{\citenamefont {{Peyrusse}}(2010)}]{PeyrusseHEDP2010}%
  \BibitemOpen
  \bibfield  {author} {\bibinfo {author} {\bibfnamefont {O.}~\bibnamefont
  {{Peyrusse}}},\ }\href {\doibase 10.1016/j.hedp.2010.09.001} {\bibfield
  {journal} {\bibinfo  {journal} {High Energy Density Physics}\ }\textbf
  {\bibinfo {volume} {6}},\ \bibinfo {pages} {357} (\bibinfo {year}
  {2010})}\BibitemShut {NoStop}%
\bibitem [{\citenamefont {Tan}\ \emph {et~al.}(2021)\citenamefont {Tan},
  \citenamefont {Kas},\ and\ \citenamefont {Rehr}}]{TanPRB2021}%
  \BibitemOpen
  \bibfield  {author} {\bibinfo {author} {\bibfnamefont {T.~S.}\ \bibnamefont
  {Tan}}, \bibinfo {author} {\bibfnamefont {J.~J.}\ \bibnamefont {Kas}}, \ and\
  \bibinfo {author} {\bibfnamefont {J.~J.}\ \bibnamefont {Rehr}},\ }\href
  {\doibase 10.1103/PhysRevB.104.035144} {\bibfield  {journal} {\bibinfo
  {journal} {Phys. Rev. B}\ }\textbf {\bibinfo {volume} {104}},\ \bibinfo
  {pages} {035144} (\bibinfo {year} {2021})}\BibitemShut {NoStop}%
\bibitem [{\citenamefont {Asato}\ \emph {et~al.}(1999)\citenamefont {Asato},
  \citenamefont {Settels}, \citenamefont {Hoshino}, \citenamefont {Asada},
  \citenamefont {Bl{\"u}gel}, \citenamefont {Zeller},\ and\ \citenamefont
  {Dederichs}}]{asato1999full}%
  \BibitemOpen
  \bibfield  {author} {\bibinfo {author} {\bibfnamefont {M.}~\bibnamefont
  {Asato}}, \bibinfo {author} {\bibfnamefont {A.}~\bibnamefont {Settels}},
  \bibinfo {author} {\bibfnamefont {T.}~\bibnamefont {Hoshino}}, \bibinfo
  {author} {\bibfnamefont {T.}~\bibnamefont {Asada}}, \bibinfo {author}
  {\bibfnamefont {S.}~\bibnamefont {Bl{\"u}gel}}, \bibinfo {author}
  {\bibfnamefont {R.}~\bibnamefont {Zeller}}, \ and\ \bibinfo {author}
  {\bibfnamefont {P.~H.}\ \bibnamefont {Dederichs}},\ }\href@noop {} {\bibfield
   {journal} {\bibinfo  {journal} {Physical Review B}\ }\textbf {\bibinfo
  {volume} {60}},\ \bibinfo {pages} {5202} (\bibinfo {year}
  {1999})}\BibitemShut {NoStop}%
\bibitem [{\citenamefont {Huhne}\ \emph {et~al.}(1998)\citenamefont {Huhne},
  \citenamefont {Zecha}, \citenamefont {Ebert}, \citenamefont {Dederichs},\
  and\ \citenamefont {Zeller}}]{huhne1998full}%
  \BibitemOpen
  \bibfield  {author} {\bibinfo {author} {\bibfnamefont {T.}~\bibnamefont
  {Huhne}}, \bibinfo {author} {\bibfnamefont {C.}~\bibnamefont {Zecha}},
  \bibinfo {author} {\bibfnamefont {H.}~\bibnamefont {Ebert}}, \bibinfo
  {author} {\bibfnamefont {P.~H.}\ \bibnamefont {Dederichs}}, \ and\ \bibinfo
  {author} {\bibfnamefont {R.}~\bibnamefont {Zeller}},\ }\href@noop {}
  {\bibfield  {journal} {\bibinfo  {journal} {Physical Review B}\ }\textbf
  {\bibinfo {volume} {58}},\ \bibinfo {pages} {10236} (\bibinfo {year}
  {1998})}\BibitemShut {NoStop}%
\bibitem [{\citenamefont {Nagayama}\ \emph {et~al.}(2019)\citenamefont
  {Nagayama}, \citenamefont {Bailey}, \citenamefont {Loisel}, \citenamefont
  {Dunham}, \citenamefont {Rochau}, \citenamefont {Blancard}, \citenamefont
  {Colgan}, \citenamefont {Coss\'e}, \citenamefont {Faussurier}, \citenamefont
  {Fontes}, \citenamefont {Gilleron}, \citenamefont {Hansen}, \citenamefont
  {Iglesias}, \citenamefont {Golovkin}, \citenamefont {Kilcrease},
  \citenamefont {Mac{F}arlane}, \citenamefont {Mancini}, \citenamefont {More},
  \citenamefont {Orban}, \citenamefont {Pain}, \citenamefont {Sherrill},\ and\
  \citenamefont {Wilson}}]{NagayamaPRL2019}%
  \BibitemOpen
  \bibfield  {author} {\bibinfo {author} {\bibfnamefont {T.}~\bibnamefont
  {Nagayama}}, \bibinfo {author} {\bibfnamefont {J.~E.}\ \bibnamefont
  {Bailey}}, \bibinfo {author} {\bibfnamefont {G.~P.}\ \bibnamefont {Loisel}},
  \bibinfo {author} {\bibfnamefont {G.~S.}\ \bibnamefont {Dunham}}, \bibinfo
  {author} {\bibfnamefont {G.~A.}\ \bibnamefont {Rochau}}, \bibinfo {author}
  {\bibfnamefont {C.}~\bibnamefont {Blancard}}, \bibinfo {author}
  {\bibfnamefont {J.}~\bibnamefont {Colgan}}, \bibinfo {author} {\bibfnamefont
  {P.}~\bibnamefont {Coss\'e}}, \bibinfo {author} {\bibfnamefont
  {G.}~\bibnamefont {Faussurier}}, \bibinfo {author} {\bibfnamefont {C.~J.}\
  \bibnamefont {Fontes}}, \bibinfo {author} {\bibfnamefont {F.}~\bibnamefont
  {Gilleron}}, \bibinfo {author} {\bibfnamefont {S.~B.}\ \bibnamefont
  {Hansen}}, \bibinfo {author} {\bibfnamefont {C.~A.}\ \bibnamefont
  {Iglesias}}, \bibinfo {author} {\bibfnamefont {I.~E.}\ \bibnamefont
  {Golovkin}}, \bibinfo {author} {\bibfnamefont {D.~P.}\ \bibnamefont
  {Kilcrease}}, \bibinfo {author} {\bibfnamefont {J.~J.}\ \bibnamefont
  {Mac{F}arlane}}, \bibinfo {author} {\bibfnamefont {R.~C.}\ \bibnamefont
  {Mancini}}, \bibinfo {author} {\bibfnamefont {R.~M.}\ \bibnamefont {More}},
  \bibinfo {author} {\bibfnamefont {C.}~\bibnamefont {Orban}}, \bibinfo
  {author} {\bibfnamefont {J.-C.}\ \bibnamefont {Pain}}, \bibinfo {author}
  {\bibfnamefont {M.~E.}\ \bibnamefont {Sherrill}}, \ and\ \bibinfo {author}
  {\bibfnamefont {B.~G.}\ \bibnamefont {Wilson}},\ }\href@noop {} {\bibfield
  {journal} {\bibinfo  {journal} {Phys. Rev. Lett.}\ }\textbf {\bibinfo
  {volume} {122}},\ \bibinfo {pages} {235001} (\bibinfo {year}
  {2019})}\BibitemShut {NoStop}%
\bibitem [{\citenamefont {{Liu}}\ \emph {et~al.}(2018)\citenamefont {{Liu}},
  \citenamefont {{Gao}}, \citenamefont {{Hou}}, \citenamefont {{Zeng}},\ and\
  \citenamefont {{Yuan}}}]{LiuNCP2018}%
  \BibitemOpen
  \bibfield  {author} {\bibinfo {author} {\bibfnamefont {P.}~\bibnamefont
  {{Liu}}}, \bibinfo {author} {\bibfnamefont {C.}~\bibnamefont {{Gao}}},
  \bibinfo {author} {\bibfnamefont {Y.}~\bibnamefont {{Hou}}}, \bibinfo
  {author} {\bibfnamefont {J.}~\bibnamefont {{Zeng}}}, \ and\ \bibinfo {author}
  {\bibfnamefont {J.}~\bibnamefont {{Yuan}}},\ }\href {\doibase
  10.1038/s42005-018-0093-5} {\bibfield  {journal} {\bibinfo  {journal}
  {Communications Physics}\ }\textbf {\bibinfo {volume} {1}},\ \bibinfo {eid}
  {95} (\bibinfo {year} {2018})}\BibitemShut {NoStop}%
\bibitem [{\citenamefont {Greenwood}(1958)}]{greenwood1958boltzmann}%
  \BibitemOpen
  \bibfield  {author} {\bibinfo {author} {\bibfnamefont {D.~A.}\ \bibnamefont
  {Greenwood}},\ }\href@noop {} {\bibfield  {journal} {\bibinfo  {journal}
  {Proceedings of the Physical Society (1958-1967)}\ }\textbf {\bibinfo
  {volume} {71}},\ \bibinfo {pages} {585} (\bibinfo {year} {1958})}\BibitemShut
  {NoStop}%
\bibitem [{\citenamefont {Kubo}(1957)}]{KuboJPSJ1957}%
  \BibitemOpen
  \bibfield  {author} {\bibinfo {author} {\bibfnamefont {R.}~\bibnamefont
  {Kubo}},\ }\href {\doibase 10.1143/JPSJ.12.570} {\bibfield  {journal}
  {\bibinfo  {journal} {Journal of the Physical Society of Japan}\ }\textbf
  {\bibinfo {volume} {12}},\ \bibinfo {pages} {570} (\bibinfo {year} {1957})},\
  \Eprint {http://arxiv.org/abs/https://doi.org/10.1143/JPSJ.12.570}
  {https://doi.org/10.1143/JPSJ.12.570} \BibitemShut {NoStop}%
\bibitem [{\citenamefont {Zeller}(2013)}]{ZellerJPCM2013}%
  \BibitemOpen
  \bibfield  {author} {\bibinfo {author} {\bibfnamefont {R.}~\bibnamefont
  {Zeller}},\ }\href {\doibase 10.1088/0953-8984/25/10/105505} {\bibfield
  {journal} {\bibinfo  {journal} {Journal of Physics: Condensed Matter}\
  }\textbf {\bibinfo {volume} {25}},\ \bibinfo {pages} {105505} (\bibinfo
  {year} {2013})}\BibitemShut {NoStop}%
\bibitem [{\citenamefont {\and William H.~Butler}(2000)}]{GonisButlerBook}%
  \BibitemOpen
  \bibfield  {author} {\bibinfo {author} {\bibfnamefont {A.~G.}\ \bibnamefont
  {\and William H.~Butler}},\ }\href@noop {} {\emph {\bibinfo {title} {Multiple
  Scattering in Solids}}}\ (\bibinfo  {publisher} {Springer},\ \bibinfo {year}
  {2000})\BibitemShut {NoStop}%
\bibitem [{\citenamefont {{Prange}}\ \emph {et~al.}(2009)\citenamefont
  {{Prange}}, \citenamefont {{Rehr}}, \citenamefont {{Rivas}}, \citenamefont
  {{Kas}},\ and\ \citenamefont {{Lawson}}}]{PrangePRB2009}%
  \BibitemOpen
  \bibfield  {author} {\bibinfo {author} {\bibfnamefont {M.~P.}\ \bibnamefont
  {{Prange}}}, \bibinfo {author} {\bibfnamefont {J.~J.}\ \bibnamefont
  {{Rehr}}}, \bibinfo {author} {\bibfnamefont {G.}~\bibnamefont {{Rivas}}},
  \bibinfo {author} {\bibfnamefont {J.~J.}\ \bibnamefont {{Kas}}}, \ and\
  \bibinfo {author} {\bibfnamefont {J.~W.}\ \bibnamefont {{Lawson}}},\ }\href
  {\doibase 10.1103/PhysRevB.80.155110} {\bibfield  {journal} {\bibinfo
  {journal} {\prb}\ }\textbf {\bibinfo {volume} {80}},\ \bibinfo {eid} {155110}
  (\bibinfo {year} {2009})},\ \Eprint {http://arxiv.org/abs/0810.3271}
  {arXiv:0810.3271 [cond-mat.mtrl-sci]} \BibitemShut {NoStop}%
\bibitem [{\citenamefont {{Alam}}\ \emph {et~al.}(2011)\citenamefont {{Alam}},
  \citenamefont {{Khan}}, \citenamefont {{Wilson}},\ and\ \citenamefont
  {{Johnson}}}]{AlamPRB2011}%
  \BibitemOpen
  \bibfield  {author} {\bibinfo {author} {\bibfnamefont {A.}~\bibnamefont
  {{Alam}}}, \bibinfo {author} {\bibfnamefont {S.~N.}\ \bibnamefont {{Khan}}},
  \bibinfo {author} {\bibfnamefont {B.~G.}\ \bibnamefont {{Wilson}}}, \ and\
  \bibinfo {author} {\bibfnamefont {D.~D.}\ \bibnamefont {{Johnson}}},\ }\href
  {\doibase 10.1103/PhysRevB.84.045105} {\bibfield  {journal} {\bibinfo
  {journal} {\prb}\ }\textbf {\bibinfo {volume} {84}},\ \bibinfo {eid} {045105}
  (\bibinfo {year} {2011})},\ \Eprint {http://arxiv.org/abs/1105.4888}
  {arXiv:1105.4888 [cond-mat.mtrl-sci]} \BibitemShut {NoStop}%
\bibitem [{\citenamefont {Starrett}\ \emph {et~al.}(2014)\citenamefont
  {Starrett}, \citenamefont {Saumon}, \citenamefont {Daligault},\ and\
  \citenamefont {Hamel}}]{StarrettPRE2014}%
  \BibitemOpen
  \bibfield  {author} {\bibinfo {author} {\bibfnamefont {C.~E.}\ \bibnamefont
  {Starrett}}, \bibinfo {author} {\bibfnamefont {D.}~\bibnamefont {Saumon}},
  \bibinfo {author} {\bibfnamefont {J.}~\bibnamefont {Daligault}}, \ and\
  \bibinfo {author} {\bibfnamefont {S.}~\bibnamefont {Hamel}},\ }\href@noop {}
  {\bibfield  {journal} {\bibinfo  {journal} {Phys. Rev. E}\ }\textbf {\bibinfo
  {volume} {90}},\ \bibinfo {pages} {033110} (\bibinfo {year}
  {2014})}\BibitemShut {NoStop}%
\bibitem [{\citenamefont {{Starrett}}\ and\ \citenamefont
  {{Saumon}}(2013)}]{StarrettPRE2013}%
  \BibitemOpen
  \bibfield  {author} {\bibinfo {author} {\bibfnamefont {C.~E.}\ \bibnamefont
  {{Starrett}}}\ and\ \bibinfo {author} {\bibfnamefont {D.}~\bibnamefont
  {{Saumon}}},\ }\href {\doibase 10.1103/PhysRevE.87.013104} {\bibfield
  {journal} {\bibinfo  {journal} {Phys. Rev. E}\ }\textbf {\bibinfo {volume}
  {87}},\ \bibinfo {eid} {013104} (\bibinfo {year} {2013})}\BibitemShut
  {NoStop}%
\bibitem [{\citenamefont {{Starrett}}\ \emph {et~al.}(2015)\citenamefont
  {{Starrett}}, \citenamefont {{Daligault}},\ and\ \citenamefont
  {{Saumon}}}]{StarrettPRE2015}%
  \BibitemOpen
  \bibfield  {author} {\bibinfo {author} {\bibfnamefont {C.~E.}\ \bibnamefont
  {{Starrett}}}, \bibinfo {author} {\bibfnamefont {J.}~\bibnamefont
  {{Daligault}}}, \ and\ \bibinfo {author} {\bibfnamefont {D.}~\bibnamefont
  {{Saumon}}},\ }\href {\doibase 10.1103/PhysRevE.91.013104} {\bibfield
  {journal} {\bibinfo  {journal} {\pre}\ }\textbf {\bibinfo {volume} {91}},\
  \bibinfo {eid} {013104} (\bibinfo {year} {2015})},\ \Eprint
  {http://arxiv.org/abs/1408.2861} {arXiv:1408.2861 [physics.plasm-ph]}
  \BibitemShut {NoStop}%
\bibitem [{\citenamefont {{Starrett}}\ \emph {et~al.}(2019)\citenamefont
  {{Starrett}}, \citenamefont {{Gill}}, \citenamefont {{Sjostrom}},\ and\
  \citenamefont {{Greeff}}}]{StarrettCPC2019}%
  \BibitemOpen
  \bibfield  {author} {\bibinfo {author} {\bibfnamefont {C.~E.}\ \bibnamefont
  {{Starrett}}}, \bibinfo {author} {\bibfnamefont {N.~M.}\ \bibnamefont
  {{Gill}}}, \bibinfo {author} {\bibfnamefont {T.}~\bibnamefont {{Sjostrom}}},
  \ and\ \bibinfo {author} {\bibfnamefont {C.~W.}\ \bibnamefont {{Greeff}}},\
  }\href {\doibase 10.1016/j.cpc.2018.10.002} {\bibfield  {journal} {\bibinfo
  {journal} {Computer Physics Communications}\ }\textbf {\bibinfo {volume}
  {235}},\ \bibinfo {pages} {50} (\bibinfo {year} {2019})},\ \Eprint
  {http://arxiv.org/abs/1804.01613} {arXiv:1804.01613 [physics.comp-ph]}
  \BibitemShut {NoStop}%
\bibitem [{\citenamefont {Karasiev}\ \emph {et~al.}(2014)\citenamefont
  {Karasiev}, \citenamefont {Sjostrom}, \citenamefont {Dufty},\ and\
  \citenamefont {Trickey}}]{KarasievPRL2014}%
  \BibitemOpen
  \bibfield  {author} {\bibinfo {author} {\bibfnamefont {V.~V.}\ \bibnamefont
  {Karasiev}}, \bibinfo {author} {\bibfnamefont {T.}~\bibnamefont {Sjostrom}},
  \bibinfo {author} {\bibfnamefont {J.}~\bibnamefont {Dufty}}, \ and\ \bibinfo
  {author} {\bibfnamefont {S.~B.}\ \bibnamefont {Trickey}},\ }\href@noop {}
  {\bibfield  {journal} {\bibinfo  {journal} {Phys. Rev. Lett.}\ }\textbf
  {\bibinfo {volume} {112}},\ \bibinfo {pages} {076403} (\bibinfo {year}
  {2014})}\BibitemShut {NoStop}%
\bibitem [{\citenamefont {{Karasiev}}\ and\ \citenamefont
  {{Hu}}(2021)}]{KarasievPRE2021}%
  \BibitemOpen
  \bibfield  {author} {\bibinfo {author} {\bibfnamefont {V.~V.}\ \bibnamefont
  {{Karasiev}}}\ and\ \bibinfo {author} {\bibfnamefont {S.~X.}\ \bibnamefont
  {{Hu}}},\ }\href {\doibase 10.1103/PhysRevE.103.033202} {\bibfield  {journal}
  {\bibinfo  {journal} {\pre}\ }\textbf {\bibinfo {volume} {103}},\ \bibinfo
  {eid} {033202} (\bibinfo {year} {2021})}\BibitemShut {NoStop}%
\bibitem [{\citenamefont {{Colgan}}\ \emph {et~al.}(2016)\citenamefont
  {{Colgan}}, \citenamefont {{Kilcrease}}, \citenamefont {{Magee}},
  \citenamefont {{Sherrill}}, \citenamefont {{Abdallah}}, \citenamefont
  {{Hakel}}, \citenamefont {{Fontes}}, \citenamefont {{Guzik}},\ and\
  \citenamefont {{Mussack}}}]{ColganApJ2016}%
  \BibitemOpen
  \bibfield  {author} {\bibinfo {author} {\bibfnamefont {J.}~\bibnamefont
  {{Colgan}}}, \bibinfo {author} {\bibfnamefont {D.~P.}\ \bibnamefont
  {{Kilcrease}}}, \bibinfo {author} {\bibfnamefont {N.~H.}\ \bibnamefont
  {{Magee}}}, \bibinfo {author} {\bibfnamefont {M.~E.}\ \bibnamefont
  {{Sherrill}}}, \bibinfo {author} {\bibfnamefont {J.}~\bibnamefont
  {{Abdallah}}, \bibfnamefont {J.}}, \bibinfo {author} {\bibfnamefont
  {P.}~\bibnamefont {{Hakel}}}, \bibinfo {author} {\bibfnamefont {C.~J.}\
  \bibnamefont {{Fontes}}}, \bibinfo {author} {\bibfnamefont {J.~A.}\
  \bibnamefont {{Guzik}}}, \ and\ \bibinfo {author} {\bibfnamefont {K.~A.}\
  \bibnamefont {{Mussack}}},\ }\href {\doibase 10.3847/0004-637X/817/2/116}
  {\bibfield  {journal} {\bibinfo  {journal} {\apj}\ }\textbf {\bibinfo
  {volume} {817}},\ \bibinfo {eid} {116} (\bibinfo {year} {2016})},\ \Eprint
  {http://arxiv.org/abs/1601.01005} {arXiv:1601.01005 [astro-ph.SR]}
  \BibitemShut {NoStop}%
\bibitem [{\citenamefont {{Faussurier}}\ and\ \citenamefont
  {{Blancard}}(2018)}]{FaussurierPRE2018}%
  \BibitemOpen
  \bibfield  {author} {\bibinfo {author} {\bibfnamefont {G.}~\bibnamefont
  {{Faussurier}}}\ and\ \bibinfo {author} {\bibfnamefont {C.}~\bibnamefont
  {{Blancard}}},\ }\href {\doibase 10.1103/PhysRevE.97.023206} {\bibfield
  {journal} {\bibinfo  {journal} {\pre}\ }\textbf {\bibinfo {volume} {97}},\
  \bibinfo {eid} {023206} (\bibinfo {year} {2018})}\BibitemShut {NoStop}%
\bibitem [{\citenamefont {{Son}}\ \emph {et~al.}(2014)\citenamefont {{Son}},
  \citenamefont {{Thiele}}, \citenamefont {{Jurek}}, \citenamefont {{Ziaja}},\
  and\ \citenamefont {{Santra}}}]{SonPRX2014}%
  \BibitemOpen
  \bibfield  {author} {\bibinfo {author} {\bibfnamefont {S.-K.}\ \bibnamefont
  {{Son}}}, \bibinfo {author} {\bibfnamefont {R.}~\bibnamefont {{Thiele}}},
  \bibinfo {author} {\bibfnamefont {Z.}~\bibnamefont {{Jurek}}}, \bibinfo
  {author} {\bibfnamefont {B.}~\bibnamefont {{Ziaja}}}, \ and\ \bibinfo
  {author} {\bibfnamefont {R.}~\bibnamefont {{Santra}}},\ }\href {\doibase
  10.1103/PhysRevX.4.031004} {\bibfield  {journal} {\bibinfo  {journal}
  {Physical Review X}\ }\textbf {\bibinfo {volume} {4}},\ \bibinfo {eid}
  {031004} (\bibinfo {year} {2014})},\ \Eprint {http://arxiv.org/abs/1404.5484}
  {arXiv:1404.5484 [physics.plasm-ph]} \BibitemShut {NoStop}%
\bibitem [{\citenamefont {{Piron}}\ and\ \citenamefont
  {{Blenski}}(2013)}]{PironHEDP2013}%
  \BibitemOpen
  \bibfield  {author} {\bibinfo {author} {\bibfnamefont {R.}~\bibnamefont
  {{Piron}}}\ and\ \bibinfo {author} {\bibfnamefont {T.}~\bibnamefont
  {{Blenski}}},\ }\href {\doibase 10.1016/j.hedp.2013.07.002} {\bibfield
  {journal} {\bibinfo  {journal} {High Energy Density Physics}\ }\textbf
  {\bibinfo {volume} {9}},\ \bibinfo {pages} {702} (\bibinfo {year}
  {2013})}\BibitemShut {NoStop}%
\bibitem [{\citenamefont {{Perrot}}(1988)}]{PerrotPA1988}%
  \BibitemOpen
  \bibfield  {author} {\bibinfo {author} {\bibfnamefont {F.}~\bibnamefont
  {{Perrot}}},\ }\href {\doibase 10.1016/0378-4371(88)90157-4} {\bibfield
  {journal} {\bibinfo  {journal} {Physica A Statistical Mechanics and its
  Applications}\ }\textbf {\bibinfo {volume} {150}},\ \bibinfo {pages} {357}
  (\bibinfo {year} {1988})}\BibitemShut {NoStop}%
\bibitem [{\citenamefont {Krief}\ \emph {et~al.}(2018)\citenamefont {Krief},
  \citenamefont {Kurzweil}, \citenamefont {Feigel},\ and\ \citenamefont
  {Gazit}}]{krief2018effect}%
  \BibitemOpen
  \bibfield  {author} {\bibinfo {author} {\bibfnamefont {M.}~\bibnamefont
  {Krief}}, \bibinfo {author} {\bibfnamefont {Y.}~\bibnamefont {Kurzweil}},
  \bibinfo {author} {\bibfnamefont {A.}~\bibnamefont {Feigel}}, \ and\ \bibinfo
  {author} {\bibfnamefont {D.}~\bibnamefont {Gazit}},\ }\href@noop {}
  {\bibfield  {journal} {\bibinfo  {journal} {The Astrophysical Journal}\
  }\textbf {\bibinfo {volume} {856}},\ \bibinfo {pages} {135} (\bibinfo {year}
  {2018})}\BibitemShut {NoStop}%
\end{thebibliography}%

\end{document}